\documentclass{aa}
\usepackage{graphicx}
\usepackage{epsf}
\usepackage{times}
\begin{document}

\def\kms{\mbox{km\,s$^{-1}$}}
\def\Hubble{\mbox{km\,s$^{-1}$\,Mpc$^{-1}$}}
\def\Doppler{\mathcal{D}}
\def\lsim{\raisebox{-.5ex}{$\;\stackrel{<}{\sim}\;$}}
\def\gsim{\raisebox{-.5ex}{$\;\stackrel{>}{\sim}\;$}}
\def\Snutspace{$(S,\nu,t)$-space}
\def\lgSnutspace{$(\lg S,\lg \nu,\lg t)$-space}
\newcommand{\mrm}[1]{\mathrm{#1}}
\newcommand{\dmrm}[1]{_{\mathrm{#1}}}
\newcommand{\umrm}[1]{^{\mathrm{#1}}}
\newcommand{\Frac}[2]{\left(\frac{#1}{#2}\right)}
\newcommand{\eqref}[1]{Eq.~(\ref{#1})}
\newcommand{\eqsref}[2]{Eqs~(\ref{#1}) and (\ref{#2})}
\newcommand{\eqssref}[2]{Eqs~(\ref{#1}) to (\ref{#2})}
\newcommand{\figref}[1]{Fig.~\ref{fig:#1}}
\newcommand{\tabref}[1]{Table~\ref{tab:#1}}
\newcommand{\secref}[1]{Sect.~\ref{sec:#1}}

\title{Simultaneous observations of the quasar 3C~273 with INTEGRAL\thanks
     {Based on observations with INTEGRAL, an ESA project with instruments 
     and science data centre funded by ESA member states (especially the PI 
     countries: Denmark, France, Germany, Italy, Switzerland, Spain), Czech 
     Republic and Poland, and with the participation of Russia and the USA.}
,  XMM-Newton \thanks{ Based on observations with {\it XMM-Newton}, an ESA science mission with instruments and contributions directly funded by ESA member states and the USA (NASA).} and  RXTE}

  \author{
T.J.-L. Courvoisier \inst{1,}\inst{2}, 
V. Beckmann \inst{1,}\inst{3},
G. Bourban \inst{2},
J. Chenevez \inst{4},
M. Chernyakova \inst{1,}\inst{2},
S. Deluit \inst{1,}\inst{2},
P. Favre \inst{1,}\inst{2},
J.E. Grindlay \inst{5},
N. Lund \inst{4},
P. O'Brien \inst{6},
K. Page \inst{6},
N. Produit \inst{1,}\inst{2},
M. T\"urler \inst{1,}\inst{2},
M.J.L. Turner \inst{6},
R. Staubert \inst{3},
M. Stuhlinger \inst{3},
R. Walter \inst{1,}\inst{2},
A.A. Zdziarski \inst{7}
}
\institute{
\textit{INTEGRAL} Science Data Centre, ch. d'\'Ecogia 16, CH-1290 Versoix, Switzerland \and
Geneva Observatory, ch. des Maillettes 51, CH-1290 Sauverny, Switzerland \and
Institut f\"ur Astronomie und Astrophysik, Universit\"at T\"ubingen, Sand 1,
D-72076 T\"ubingen, Germany \and
Danish Space Research Institute, Juliane Maries Vej 30, DK-2100 Copenhagen Ø, Denmark \and
Harvard-Smithsonian Center for Astrophysics, 
60 Garden St., Cambridge, MA 02138, U.S.A \and
Department of Physics and Astronomy, University of Leicester, University road, Leicester LE1 7RH, United Kingdom \and
N. Copernicus Astronomical Ctr, Bartycka 18, PL--00716 Warsaw, Poland
}
\offprints{T. Courvoisier (ISDC)}
\mail{Thierry.Courvoisier@obs.unige.ch}
\date{Received /Accepted }

\abstract{
{\it INTEGRAL}  has observed the bright quasar 3C~273 on 3 epochs in January 2003 as one of the first observations of the open programme. The observation on January 5 was simultaneous with {\it RXTE} and {\it XMM-Newton} observations. We present here a first analysis of the continuum emission as observed by these 3 satellites in the band from $\simeq$\,3\,keV to $\simeq$\,500\,keV. The continuum spectral energy distribution of 3C~273 was observed to be weak and steep in the high energies during this campaign. We present the actual status of the cross calibrations between the instruments on the three platforms using the  calibrations available in June 2003.
}

\authorrunning{Courvoisier et al.}
\titlerunning{{\it INTEGRAL} Observations of 3C~273}
\maketitle

\section{Introduction}

\begin{figure}[b]
\includegraphics[width=9cm]{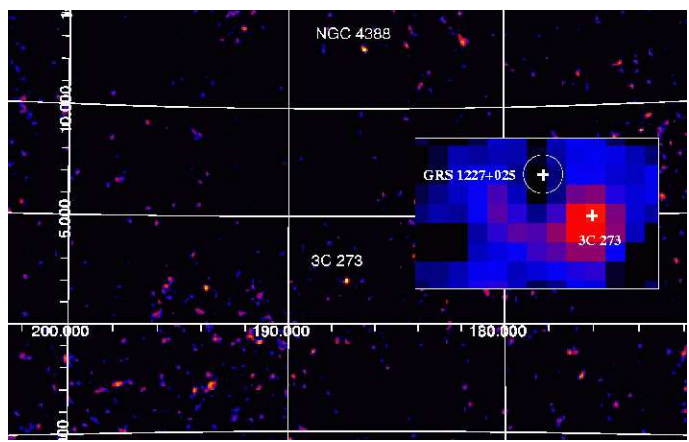}
\hfill
\caption{\label{fig:geometry}%
 IBIS/ISGRI significance image in the 25-40\,keV energy band of the field of 3C~273 and NGC~4388. The insert shows a 25-40\,keV flux image of the area surrounding 3C~273 and where GRS~1227+025 would be expected. The 3C~273 count rate is about 2.7 counts\,s$^{-1}$, that of GRS~1227+025 is less than 0.4\,counts\,s$^{-1}$.}
\end{figure}

{\it INTEGRAL} (\cite{winkleretal2003}) observed the bright quasar 3C~273 (see \cite{courvoisier98} for a review of the properties of the quasar) at 3 epochs in January 2003.
This is the beginning of a programme meant to monitor the high energy emission of the object in order to measure the different components contributing to the emission above 1 keV and their respective variability. Some observations of the {\it INTEGRAL} programme are conducted in coordination with the {\it XMM-Newton} and {\it RXTE} satellites in order to provide cross calibrations between the instruments on board these three platforms.

We report here a first analysis of the {\it INTEGRAL} data together with the results of the {\it XMM-Newton} and {\it RXTE} observations that were conducted simultaneously with the {\it INTEGRAL} measurements in January 2003. The {\it INTEGRAL} data span 12 days while the {\it XMM-Newton} and {\it RXTE} data were obtained on a single date during  the first {\it INTEGRAL} observations.

\section{INTEGRAL observations}

{\it INTEGRAL}  observed 3C~273 during  revolution 28 starting January 5, 2003 for $1.2 \times 10^5$\,s, during  revolution 30 starting January 11, 2003 for $10^4$\,s and during  revolution 32 starting January 17, 2003 for $1.1 \times 10^5$\,s. The observations were performed using the 25 points square dithering pattern that is expected to give the best results for the SPI analysis except for 23\,000\,s during which the satellite pointed stably on the source. Only the revolution 28 {\it INTEGRAL} observation was strictly simultaneous with the {\it RXTE} and {\it XMM-Newton} observations. There is as of now no evidence of significant 3C~273 flux variations observed by the  {\it INTEGRAL} instruments between the revolutions 28, 30 and 32. We therefore combine here all available data. Future analysis will be done to quantify possible flux  variability.

\subsection{IBIS/ISGRI results}

We present in Fig. 1 a significance image of the IBIS/ISGRI (\cite{lebrunetal2003})  25-40\,keV data accumulated over $1.62 \times 10^5$\,s. We used version 2.0 of ISDC's (\cite{courvoisieretal2003}) Offline Science Analysis (OSA) software. The algorithms used in the analysis are described in \cite{goldwurmetal2003}. 3C~273 and the bright Seyfert 2 galaxy NGC~4388 are clearly detected at a significance level of 10.5 for 3C~273  and 9.3 for NGC~4388. No source is detected in the surrounding of 3C~273. Other maxima in the significance image will be analysed to assert their nature as either noise or possible sources below the present level of detection. The 3C~273 countrate in the 25 to 40\,keV band is $2.66 \pm 0.12 $\,counts\,s$^{-1}$ (here and in the following we use $1\,\sigma$ confidence level when quoting uncertainties). The flux of 3C~273 at $50$\,keV is about $7$\,mCrab. 

We have used the image shown in Fig. 1 to assess the reality of the existence of the source GRS1227+025 (\cite{jourdainetal1992}) which was claimed to be detected during a 1990 {\it SIGMA} observation. We show in the insert  of Fig. 1 a flux image of the region near 3C~273 with the  GRS~1227+025 possible localisation marked. GRS1227+025 was found in the 1990 {\it SIGMA} data to be much brighter than 3C~273 which was not seen during that observation.  \cite{leach1996} have searched for a possible counterpart to GRS1227+025 in the {\it ROSAT} data and concluded that there is no likely candidate and that the source, if real, must be transient or highly absorbed. Our data show no source at the position of GRS 1227+025 (Fig. 1). At the present level of the analysis we can conclude that a conservative upper limit to the flux of GRS1227+025 is 1/2 that of 3C~273 in the 25-40\,keV band. This estimate is based on our understanding of the systematic uncertainties rather than on statistics. We  expect that more stringent limits will be obtained in the future. The other source in the field of view that is seen in Fig. 1 is the bright Seyfert 2 galaxy NGC~4388. This source is $10.6^{\circ}$ away and does not contribute therefore  to the fluxes given by the instruments used in this study.

We extracted the ISGRI spectrum of 3C~273 accumulated over $1.62 \times 10^5$\,s using the data from the  revolutions 28, 30 and 32.  We used an intermediate version of the  OSA software and the matrices describing the instrument as available at the ISDC  in June 2003. We have ignored the channels below 25\,keV and above 100\,keV. The resulting count spectrum can be well fitted with a single power law function between 25 and 100\,keV. The photon index is $1.95 \pm 0.2$.  

\begin{figure}[b]
\centering
\includegraphics[width=8cm]{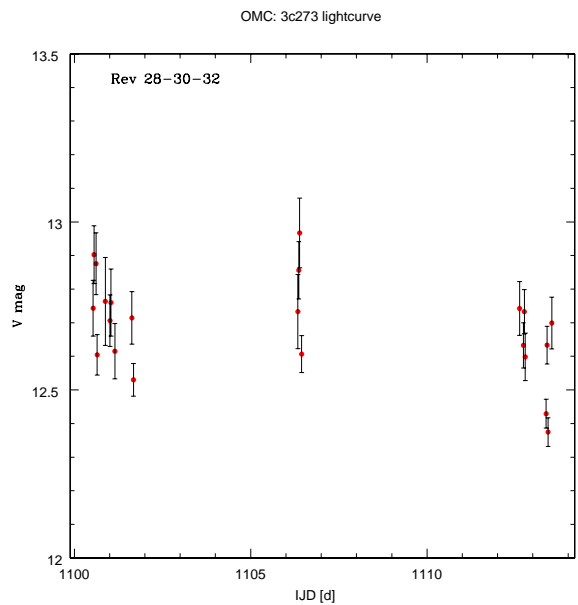}
\hfill
\caption{\label{fig:geometry}%
 OMC light curve. IJD is the fractional number of days since January 1, 2000 at 0\,UT}
\end{figure}

\subsection{SPI results}

The analysis of the SPI (\cite{vedrenneetal2003}) data is based on 74 dithering pointings with a
total exposure time of 147 ksec. From the total of 83 dithering
pointings which were taken during the 3C~273 observation, 9 had to be
excluded from the SPI analysis as they either were affected by strong
solar activity or had been influenced by the radiation belts. As the
SPI data are background dominated, a careful background substraction
is essential in order to get reasonable results, especially for weak
sources. A time dependent background model has been applied to the
data, based on the saturated events seen by the detector.
The image reconstruction used for the analysis presented
here is done by using the Iterative Removal Of
Sources (IROS) method (\cite{hammersleyetal1992}) which is implemented in the SPIROS
software (\cite{skinneretal2003}).
To get precise flux values, the source positions of the
three brightest sources in the field (3C~273, 3C~279, and NGC~4388) have
been fixed to their catalogue values, and fluxes have been extracted.
No other source with a significance larger than $ 3 \sigma$ has been detected
in the SPI data.
For extraction of a rough spectrum, five  energy bins (with boundaries at $20,
40, 100, 200, 500, 1000 \, \rm keV$) have been applied to the data.
3C~273 is detectable in the SPI data up to at least 200 keV. The flux
value in the $200 - 500 \, \rm keV$ band is  $f = (5.5 \pm 4.9) \times 10^{-4} \rm \,
photons \, cm^{-2} \, sec^{-1}$. Simulations of weak sources have
also shown that the analysis software tends to overestimate their flux at high ($> 200 \rm \, keV$) energies, and thus a
detection up to $500 \rm \, keV$ cannot yet be confirmed by the SPI data.
The SPI spectrum is consistent with a single power law with photon
index $\Gamma = 1.66 \pm 0.28$ and the flux averaged over the whole
observation is $\simeq 10.0 \pm 2.5 \, \rm mCrab$.

\subsection{JEM-X results}

We analysed all the data in which 3C~273 was on-axis, this amounts to 32\,000\,s of JEM-X 2 (\cite{lundetal2003}) data (the time during which {\it INTEGRAL} was in staring mode plus those pointings where the source was on-axis).  The analysis was performed by fixing the position of the source, extracting the counts spectra for each pointing individually and adding the results weighted by the exposure time of the individual pointings. 

The spectral extraction is based on an algorithm which is similar
to photon tagging (\cite{fenimore1986}). Each
detected photon is backprojected through the mask and its
contribution to the source and background flux estimates
for the particular source is computed. The spectrum is built
up by the net contribution from all detected photons. The implementation of the algorithm in the JEM-X case is described in \cite{westergaardetal2003}. 

When performing a single power law spectral analysis of the extracted spectrum we obtain a slope $\Gamma = 1.6 \pm 0.23$. The 3-10\,keV flux is $(4.0 \pm 0.2) \times 10^{-11}$\,ergs\,cm$^{-2}$\,s$^{-1}$.

\subsection{OMC light curve}

The Optical Monitoring Camera (OMC, \cite{mashesseetal2003}) provides a
simultaneous set of observations of several sources in the  {\it INTEGRAL} field of
view. We considered all exposures of 100\,s of 3C~273 and extracted
square boxes (187\,$\times$\,187\,arcsec$^2$) centered on 3C~273. The images of these
boxes have been combined in a single averaged image per science window
and their analysis was then performed by ``square photometry''
using the IRAF image package IMSTATISTICS, first on the whole box and
then on small boxes centered on sources close to 3C~273 in order to
remove their contribution to the flux of the quasar and to the
estimate of the background. Indeed as the pixel size of OMC CCD is $\simeq$\,17
arcsec and severly undersamples the point spread function (PSF),  a circular aperture photometry or  PSF fit are
inapropriate.

The resulting light curve (Fig. 2) shows an average   flux level of m$_V = 12.72 \pm 0.02 $ during revolution 28, m$_V = 12.79 \pm 0.03 $ during revolution 30 and m$_V = 12.61 \pm 0.03 $ during revolution  32.  Figure 2 shows some level of variation ($\simeq 0.3$ magnitude) during all revolutions. A careful examination of the data  gives no indication that  instrumental effects   could affect these measurements. In particular, these variations are still present when the light from the whole window   is considered. Furthermore, the 2 other  sources in the window are too weak (V = 13.2 and V = 14.9) to account for the observed variations. No source on the outside of the box is expected to influence the total flux measured, the background was not observed to vary significantly. Further work is in progress to understand these variations.

\begin{figure*}
\epsfxsize=15.0cm
\epsfbox{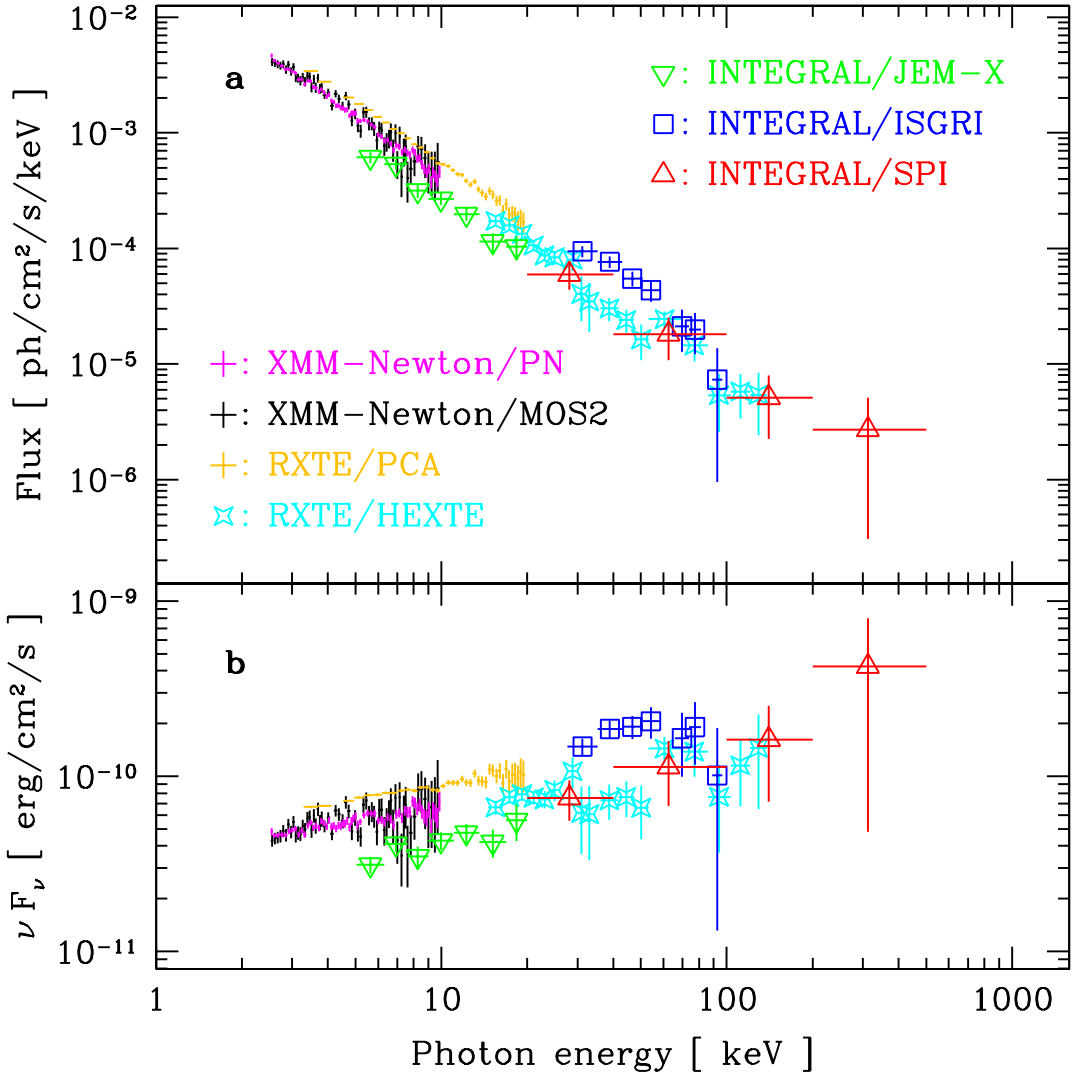}
\caption[]{\label{fig:spectrum}
The spectral energy distributions measured 
by the 3 high energy instruments of INTEGRAL, by XMM-Newton and RXTE.}
\end{figure*}

\begin{figure*}
\epsfxsize=15.0cm
\epsfbox{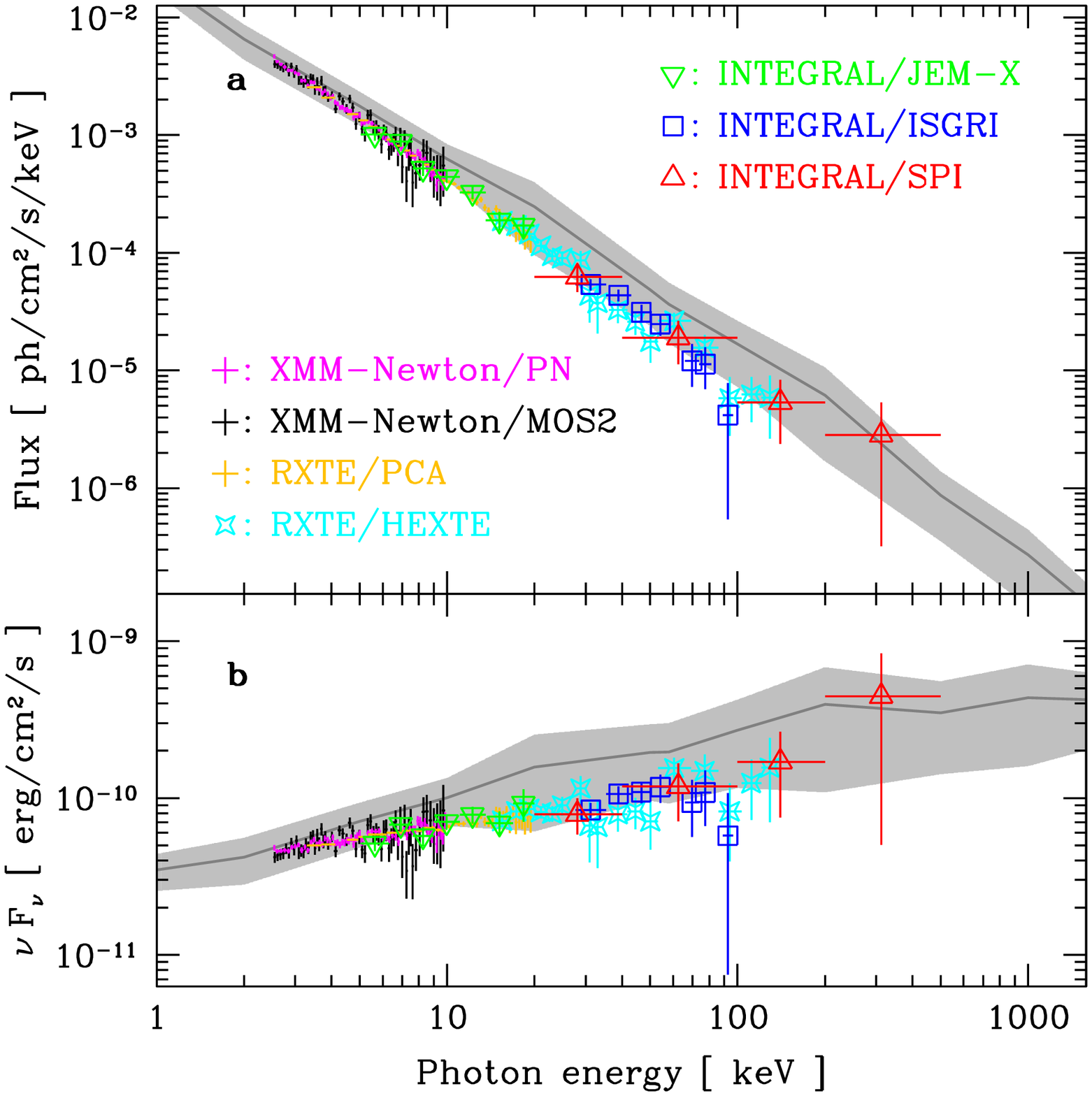}
\caption[]{\label{fig:spectrum}
The spectral energy distributions of figure 3 normalised to the PN camera of {\it XMM-Newton} flux and compared with the historic average (continuous line) and observed range of variations as given in \cite{turleretal1999}}
\end{figure*}

\section{XMM-Newton Observations}

{\it XMM-Newton} observed 3C~273 January 5, 2003 from 14:22 to 16:45 and from 17:29 to 18:53 (UT).

From the EPIC MOS/PN cross-calibration\footnote{XMM-SOC-CAL-TN-0018,
2003-04-04} it is known that for the Small
Window (SW) mode the spectral shape obtained with  MOS2 and PN are consistent with one
another. The MOS1 camera was in several different modes during the observations. The MOS2 camera was in the Small Window mode throughout the observation, its data is therefore retained for the remainder of this analysis.

The MOS2 Small Window  observations show pile-up in the region of the source center.
Therefore the central region of the image is excluded from the
analysis. To examine the source area containing pile-up we
excluded a circular region of the source center and decreased the radii of
this regions until pile-up effects become perceptible.
This is verified by the comparison of the pattern distribution and pattern
fraction of valid patterns of type single, double, triple, and quadruple
with model curves using the SAS V.5.4.1 task EPATPLOT V.1.1.8.

The PN observations do not show pile-up effects. The small discrepancy
seen in the pattern distribution below 600\,eV is related to a PN
calibration issue and not related to pile-up.

For the analysis presented in this paper circular source regions are used
with size of 0--40\,arcsec for PN SW and 10--40\,arcsec
for MOS2 SW. The spectra are generated out of single events (pattern 0)
only.
The PN background was selected from two boxes near the border of the small
window chip range. For MOS2 it is impossible to define a background region
on
CCD1 because the PSF fills completely the small window area. The
background
is determined from two circular regions of CCD3/6  where no background
source is identifiable. 

We performed a spectral fit above 3\,keV, thus avoiding the complex soft X-ray excess region, for the first period of observation.  We  used  a simple power law and obtain a slope $\Gamma = 1.74 \pm 0.03$ (PN) and $1.73 \pm 0.08$ (MOS2). The 3-10\,keV fluxes are $(6.8 \pm 0.7) \times 10^{-11}$\,ergs\,cm$^{-2}$\,s$^{-1}$ (PN) and  $(7.0 \pm 1.8) \times 10^{-11}$\,ergs\,cm$^{-2}$\,s$^{-1}$ (MOS2). The flux uncertainties take  the uncertainty on the slope and the normalisation into account. These results are entirely consistent with those obtained during the second period of observation.

There is no evidence of an Fe K$_\alpha$ line. Upper limits for the equivalent width of a neutral or ionised Fe line are $< 52$\,eV for a narrow ($\sigma = 0.01$\,keV) and $< 90$\,eV for a broad ($\sigma = 0.5$\,keV) line.

\section{RXTE Observations}

{\it RXTE}  observed 3C~273 January 5, 2003 between 14:22 and 19:00 (UT). 

For {\it RXTE} only PCU0 and PCU2 were enabled. Because the count rate of the
quasar observation was very low only the top layers were taken into
account
to  maximise the signal to noise ratio. Since May 2000  the top layer of
PCU0 is damaged, so only the top layer of PCU2 was used for spectral
analysis.

The PCA Standard Mode 2 data were reduced using FTOOLS V5.2. The response
file was generated using PCARMF V8.0. The background
estimation was done with PCABACKEST V3.0 using the L7-240 background model
recommended for faint sources by the NASA/GSFC RXTE guest observer facility.

Both HEXTE clusters were combined to get higher signal to noise
ratio. Nevertheless the count rate is very weak. The poor statistics causes PCA/HEXTE cross-calibration
problems. Fitting the PCA and HEXTE (20-190 keV) data simultaneously the
fit is completely dominated by the PCA, the resulting $\Gamma$(PCA/HEXTE)
is identical to $\Gamma$(PCA). We show here the HEXTE data over the complete energy range, aware of its low significance, to show how it compares with the other instruments.

The results of the PCA spectral analysis give  $\Gamma = 1.73 \pm 0.02$ and a 3-10\,keV flux of  $(9.2 \pm 0.5) \times 10^{-11}$\,ergs\,cm$^{-2}$\,s$^{-1}$. There is no evidence for an Fe line. 

\begin{table}[b]
\caption[]{Intercalibration factors deduced from a fit to the 3C~273  high-energy data with a power law.}
\begin{tabular}{rrrr} \hline \hline
XMM-PN & XMM-MOS2  & RXTE-PCA & RXTE-HEXTE   \\ \hline 
1.0 (fix) & $1.024 \pm0.025$ & $1.34\pm 0.02$  & $0.93 \pm 0.06$ \\ \hline \hline JEM-X & IBIS-ISGRI & SPI \\ \hline $0.61 \pm0.05$ & $1.76 \pm 0.2 $& $0.95 \pm 0.3$\\ \hline
\end{tabular}
\end{table}

\section{Discussion}

We present in Fig. 3 the spectral energy distributions obtained by the high energy instruments of {\it INTEGRAL} and by {\it XMM-Newton} and by {\it RXTE}. The photon distributions as well as  a $\nu\,f_{\nu}$ representation are given. The data are shown   without taking into account the factors that correct the normalisations obtained with the different instruments. It is clear from Fig. 3 and Table 1 that the measurement of the detailed shape of the emission over large spectral ranges spanning more than one instrument as it is needed to establish the presence of reflection components and cut-offs hinges on the  intercalibrations between the instruments not only on different satellites but also on the same platform.

The average V magnitude of 3C~273 quoted in \cite{courvoisier98} is 12.9\,mag. The magnitudes measured here are between 12.61\,mag and 12.79\,mag showing that the object was slightly brighter in the optical domain during this campaign than it usually is. 

The high energy continuum spectral energy distribution measured here (Figure 4) using cross calibration factors of the instruments given in Table 1 (i.e. aligning all instruments to the PN Camera of {\it XMM-Newton})   is well represented by a power law with a photon index of $1.73 \pm 0.015$  and 
normalisation of $(2.24 \pm 0.05) \times 10^{-2}$ photons\,cm$^{-2}$\,s$^{-1}$\,keV$^{-1}$ at 1 keV. The reduced chi-square is 0.94 for 292 degrees of freedom. This measurement  differs from the spectrum of \cite{lichtietal95}  which has a similar normalisation at 1 keV  ($(2.33 \pm 0.04) \times 10^{-2}$ photons\,cm$^{-2}$\,s$^{-1}$\,keV$^{-1}$) but is flatter, the slope being $1.6 \pm 0.01$. 

The continuum spectral energy distribution shown in Fig. 4 is normalised to the PN camera of {\it XMM-Newton} flux value.  It is slightly below the historic average of \cite{turleretal1999} in the low energies. The slope deduced from the measurements presented here is steeper than that suggested by the historical average which is close to that of \cite{lichtietal95}. The continuum spectral energy distribution between 10 and 100 keV is therefore close to the weakest observed to date (Fig.4).

We conclude that there is an important cross calibration effort to be performed  in order to be able to confidently use multi-instrument data to model the emission of sources over several decades of energy. Fig. 3 indicates that there is a significant mismatch in the present calibrations of ISGRI and SPI (June 2003). The normalisation of 3C~273 deduced with SPI (Table 1) agrees better with that obtained from {\it XMM-Newton} than  the normalisation obtained with ISGRI. The relative normalisations obtained with both {\it RXTE} instruments indicates that some cross calibrations issues remain to be solved. The extensive SPI ground calibration campaign (\cite{attieetal2003}) will prove very important in settling the issue of cross-calibration in the near future.

3C~273 has been weak in the high energy domain in January 2003 and the spectral slope is steeper than in previous campaigns. There is no evidence for components other than a single power-law above 3\,keV in the data presented here.

\begin{acknowledgements}
We are grateful to a large collaboration which contributes to the multi-wavelength studies of 3C~273 for their contributions to this project.
\end{acknowledgements}

\end{document}